# ESTIMATING PREDICTABILITY: REDUNDANCY AND SURROGATE DATA METHOD*

M. Paluš[†], L. Pecen and D. Pivka

*Institute of Computer Science, Academy of Sciences of the Czech Republic*
*Pod vodárenskou věží 2, 182 07 Prague 8, Czech Republic*

June 20, 1995

**Abstract**

A method for estimating theoretical predictability of time series is presented, based on information-theoretic functionals—redundancies and surrogate data technique. The redundancy, designed for a chosen model and a prediction horizon, evaluates amount of information between a model input (e.g., lagged versions of the series) and a model output (i.e., a series lagged by the prediction horizon from the model input) in number of bits. This value, however, is influenced by a method and precision of redundancy estimation and therefore it is a) normalized by maximum possible redundancy (given by the precision used), and b) compared to the redundancies obtained from two types of the surrogate data in order to obtain reliable classification of a series as either unpredictable or predictable. The type of predictability (linear or nonlinear) and its level can be further evaluated. The method is demonstrated using a numerically generated time series as well as high-frequency foreign exchange data and the theoretical predictability is compared to performance of a nonlinear predictor.

## 1 Introduction

Forecasting short-term evolution of a system based on measurements of its past history, in particular, time series prediction, is a challenging task from both theoretical and practical points of view. A number of new prediction methods have been developed recently based on ideas from nonlinear dynamics and theory of deterministic chaos [2, 28], in addition to traditional (mostly linear) methods [23]. They are being employed in a broad area of scientific disciplines and application areas. Particularly, forecasting in economics attracts specialists from various fields of mathematical, physical and computer sciences.

In a prediction problem, one can train various types of predictors (i.e., fit considered models) and then evaluate their performance by computing prediction errors. Failure of some predictors does not necessarily mean that a series is principally unpredictable. Thus, a method for evaluating theoretical predictability of a series is desirable. Results of various methods for time series analysis have more or less direct implications for predictability of the series. Different methods, however, evaluate different aspects of dynamics of a series and by simple summarizing the results from different methods a clear picture about dependence structures in a series can be hardly made. Moreover, a number of methods for evaluating nonlinear aspects of time series can be influenced by numerical artifacts and spurious effects caused by linear properties of data under study (see [20] and references within).

---

In this paper we present a nonparametric method for estimating theoretical predictability of time series, based on information-theoretic functionals—redundancies and the surrogate data technique. The redundancy, that can be designed for a chosen model and a prediction horizon, evaluates the amount of information between the model input (e.g., the lagged versions of the series) and the model output (i.e., the series lagged by the prediction horizon from the model input) in number of bits. This value, however, is influenced by a method and precision of redundancy estimation and therefore it is normalized by the maximum possible redundancy, which is related to the entropy of the series. The resulted value is the theoretical predictability in per cents of the maximum predictability (i.e., a one-to-one mapping considering given precision). In order to avoid spurious results from possible systematic deviation of estimated values from correct ones, the same redundancy as from the raw series is also computed from two types of surrogate data sets, representing two types of concurrent null hypotheses: a) "scrambled surrogates"—representing white noise or no predictability; b) "isospectral surrogates"—representing the null hypothesis of colored noise (an autocorrelated linear stochastic process), i.e., non-zero predictability caused by linear autocorrelations. Differences (in redundancies) between the investigated data and the surrogates are evaluated statistically. As a result, the method gives one of the following qualitative characterizations of the series under study:

- The data is unpredictable (white noise);
- the data is predictable by a linear model;
- the data is predictable by a nonlinear model.

In addition, the quantitative measure of predictability can be used for comparison of different datasets or segments of a series and for finding a relation between this predictability measure and performance of predictors in use. In economic applications the method can be used for designing investment strategy and choosing a predictor appropriate for current market dynamics. The computational cost of this method is smaller than training/testing any predictor.

The theoretical predictability measures, proposed in this paper, are computed from information-theoretic functionals—redundancies, which, together with entropies and mutual information are introduced in Sec. 2. Further details can be found in [7, 10, 11, 3, 24] and references therein. The concept of surrogate data is briefly reviewed in Sec. 3. The details of the proposed method are described in Sec. 4 and possible applications of the method are demonstrated in Sec. 5 using a numerically generated time series, as well as high-frequency foreign exchange data, and the theoretical predictability is compared to performance of a nonlinear predictor.

## 2 Entropy, Information and Redundancy

Let $X$ be a discrete random variable with a set of values ("alphabet") $\Xi$ and probability mass function $p(x) = \Pr\{X = x\}$, $x \in \Xi$. We denote the probability mass function by $p(x)$, rather than $p_X(x)$, for convenience.

The *entropy* $H(X)$ of a discrete random variable $X$ is defined by

$$H(X) = -\sum_{x \in \Xi} p(x) \log p(x). \tag{1}$$

For a pair of discrete random variables $X$ and $Y$ with a joint distribution $p(x, y)$ the *joint entropy* $H(X, Y)$ is defined as

$$H(X, Y) = -\sum_{x \in \Xi} \sum_{y \in \Upsilon} p(x, y) \log p(x, y). \tag{2}$$

The *conditional entropy* $H(Y|X)$ of $Y$ given $X$ is defined as

$$H(Y|X) = \sum_{x \in \Xi} p(x) H(Y|X = x)$$

$$= -\sum_{x\in\Xi} p(x) \sum_{y\in\Upsilon} p(y|x) \log p(y|x) \qquad (3)$$

$$= -\sum_{x\in\Xi} \sum_{y\in\Upsilon} p(x,y) \log p(y|x).$$

The average amount of common information, contained in the variables $X$ and $Y$, is quantified by the *mutual information* $I(X;Y)$, defined as

$$I(X;Y) = H(X) + H(Y) - H(X,Y). \qquad (4)$$

The joint entropy of $n$ variables $X_1,\ldots,X_n$ with the joint distribution $p(x_1,\ldots,x_n)$ is defined as

$$H(X_1,\ldots,X_n) = -\sum_{x_1\in\Xi_1} \ldots \sum_{x_n\in\Xi_n} p(x_1,\ldots,x_n) \log p(x_1,\ldots,x_n). \qquad (5)$$

*Redundancy* $R(X_1;\ldots;X_n)$ quantifies the average amount of common information contained in the $n$ variables $X_1,\ldots,X_n$ and can be defined as straightforward generalization of (4):

$$R(X_1;\ldots;X_n) = H(X_1) + \ldots + H(X_n) - H(X_1,\ldots,X_n). \qquad (6)$$

Besides (6), the *marginal redundancy* $\varrho(X_1,\ldots,X_{n-1};X_n)$, quantifying the average amount of information about the variable $X_n$ contained in the variables $X_1,\ldots,X_{n-1}$, can be defined as

$$\varrho(X_1,\ldots,X_{n-1};X_n) = H(X_1,\ldots,X_{n-1}) + H(X_n) - H(X_1,\ldots,X_n). \qquad (7)$$

The relation

$$\varrho(X_1,\ldots,X_{n-1};X_n) = R(X_1;\ldots;X_n) - R(X_1;\ldots;X_{n-1}) \qquad (8)$$

can be derived by simple manipulation.

Except of the redundancy (6) and the marginal redundancy (7) we can define various types of redundancies quantifying the average amounts of information between/among variables or groups of variables. For instance, considering variables $X_1,\ldots,X_n, Y_1,\ldots,Y_m, Z_1,\ldots,Z_k$, the redundancy among the three groups of $X$'s, $Y$'s and $Z$'s is

$$R(X_1,\ldots,X_n;Y_1,\ldots,Y_m;Z_1,\ldots,Z_k) = H(X_1,\ldots,X_n) + H(Y_1,\ldots,Y_m)$$
$$+ H(Z_1,\ldots,Z_k) - H(X_1,\ldots,X_n,Y_1,\ldots,Y_m,Z_1,\ldots,Z_k). \qquad (9)$$

Or, the redundancy between the $X$'s and $Y$'s (considered together) and the $Z$'s (a generalization of the marginal redundancy for groups of variables) is

$$R(X_1,\ldots,X_n,Y_1,\ldots,Y_m;Z_1,\ldots,Z_k) = H(X_1,\ldots,X_n,Y_1,\ldots,Y_m)$$
$$+ H(Z_1,\ldots,Z_k) - H(X_1,\ldots,X_n,Y_1,\ldots,Y_m,Z_1,\ldots,Z_k). \qquad (10)$$

Also, conditional redundancies can be defined. For instance, the conditional redundancy between the groups of the $X$'s and the $Y$'s given the $Z$'s is

$$R(X_1,\ldots,X_n;Y_1,\ldots,Y_m|Z_1,\ldots,Z_k) = H(X_1,\ldots,X_n|Z_1,\ldots,Z_k)$$
$$+ H(Y_1,\ldots,Y_m|Z_1,\ldots,Z_k) - H(X_1,\ldots,X_n,Y_1,\ldots,Y_m|Z_1,\ldots,Z_k). \qquad (11)$$

Or, the conditional redundancy among $X_1,\ldots,X_n$ given $Y_1,\ldots,Y_m$ is

$$R(X_1;\ldots;X_n|Y_1,\ldots,Y_m) = H(X_1|Y_1,\ldots,Y_m) + H(X_2|Y_1,\ldots,Y_m) + \ldots$$

$$+ H(X_n|Y_1,\ldots,Y_m) - H(X_1,\ldots,X_n|Y_1,\ldots,Y_m). \tag{12}$$

Now, let the $n$ variables $X_1,\ldots, X_n$ have zero means, unit variances and correlation matrix **C**. Then, we define the *linear redundancy* $L(X_1;\ldots;X_n)$ of $X_1, X_2, \ldots, X_n$ as

$$L(X_1;\ldots;X_n) = -\frac{1}{2} \sum_{i=1}^{n} \log(\sigma_i), \tag{13}$$

where $\sigma_i$ are the eigenvalues of the $n \times n$ correlation matrix **C**.

If $X_1,\ldots,X_n$ have an $n$-dimensional Gaussian distribution, then $L(X_1;\ldots;X_n)$ and $R(X_1;\ldots;X_n)$ are theoretically equivalent [15].

Based on (8) we define the *linear marginal redundancy* $\lambda(X_1,\ldots,X_{n-1};X_n)$, quantifying linear dependence of $X_n$ on $X_1,\ldots,X_{n-1}$, as

$$\lambda(X_1,\ldots,X_{n-1};X_n) = L(X_1;\ldots;X_n) - L(X_1;\ldots;X_{n-1}). \tag{14}$$

Similarly, for any kind of the general (nonlinear) redundancy its linear equivalent exists—it can be either obtained as a combination of the redundancies of the type (14) based on related relation to the redundancies of the type (6)[1], or derived directly using relevant expressions for (multidimensional) Gaussian distributions.

The general redundancies $R$ detect all dependences in data under study, while the redundancies $L$ are sensitive only to linear structures. (For detailed discussion see [20].)

## 3 Surrogate Data

The surrogate data method, related to the technique of bootstrap [5, 6], has been methodologically introduced in nonlinear dynamics by Theiler et al. [25, 26] as a method for testing nonlinearity. The basic idea in the surrogate-data based nonlinearity test is to compute a *nonlinear* statistic for data under study and for an ensemble od realizations of a linear stochastic process, which mimics "linear properties" of the studied data. If the computed statistic for the original data is significantly different from the values obtained for the surrogate set, one can infer that the data were not generated by a linear process; otherwise the null hypothesis, that a linear model fully explains the data, is accepted, and the data can be further analyzed and predicted using well-developed linear methods.

In general, surrogate data are artificially generated data, which mimic statistical properties of the data under study, but not the property which is tested for. If *any* temporal dependence (predictability) is under question, a test can use the null hypothesis of an independent identically distributed (iid) process (white noise) and so called *scrambled* surrogates are used: The original series is mixed in temporal order, so that all original temporal dependences (if any) are eliminated in the scrambled surrogates, but the mean, the variance and the histogram of the original data are preserved. In the case of testing for nonlinearity, the surrogate data should have the same spectrum[2] and, consequently, the autocorrelation function ("linear properties") as the original data under study, however, surrogate data are generated as realizations of a linear stochastic process. It can be achieved in the following way: Compute the Fourier transform (FT) of the original data, randomize the phases but keep the original absolute values of the Fourier coefficients (i.e., the spectrum) and perform the inverse FT into the time domain. The resulting time series is a realization of a linear stochastic process with the same spectrum as the original data.

Another way to generate a linear stochastic surrogate is fitting an ARMA (auto-regressive moving-average) model. Theiler et al. [25] discuss relations between the FT- and ARMA-based surrogates and argue that, for testing hypotheses, the FT-based surrogates are better. Theiler and Prichard [27] demonstrate that a test for nonlinearity based on the FT surrogates can be more powerful than the same test based on the ARMA surrogates, and an actual nonlinearity in a data can be neglected by a test using the latter. In the following we will consider the FT surrogates.

---

[1] I.e., any type of the redundancy can be expressed as a combination of the redundancies of the type (6). Such an expression always contains the redundancy among all the variables under study, as far as an n-dimensional redundancy generally cannot be reduced to a sum of lower-dimensional redundancies.

[2] Also, preservation of histogram is usually required. A histogram transformations used for this purpose is described in [20] and references within.

## 4 The Method

In Section 2 we have introduced various types of redundancies in order to demonstrate that various kinds of dependence structures (models) can be studied and tested, including models employing multivariate time series[3]. In the following we will focus our attention on a typical situation of a univariate time series $\{x(t)\}$, considered as a realization of a stationary and ergodic stochastic process $\{X(t)\}$. Considering a prediction model of order $n$, an information theoretic functional of interest is the marginal redundancy $\varrho(X(t), X(t+\tau_1), \ldots, X(t+\tau_{n-1}); X(t+\tau_n))$, which evaluates the amount of information, contained in the $n$ variables $X(t), X(t+\tau_1), \ldots, X(t+\tau_{n-1})$, about the variable $X(t+\tau_n)$. In other words, the marginal redundancy $\varrho(X(t), X(t+\tau_1), \ldots, X(t+\tau_{n-1}); X(t+\tau_n))$ quantifies the dependence of $X(t+\tau_n)$ on $X(t), X(t+\tau_1), \ldots, X(t+\tau_{n-1})$ and thus it quantifies the theoretical predictability of $X(t+\tau_n)$ (the output of a model) based on the $n$ variables $X(t), X(t+\tau_1), \ldots, X(t+\tau_{n-1})$ (the input of a model). The redundancy, however, does not specify the relation between $X(t+\tau_n)$ and $X(t), X(t+\tau_1), \ldots, X(t+\tau_{n-1})$. It only indicates that such a relation exists, and, as we show below, it specifies how strong or weak it is and whether it is linear or nonlinear.

Considering ergodicity of an underlying process, all information-theoretic functionals can be estimated using time averages instead of ensemble averages; in particular, correlation matrices in (13) are obtained as the time averages over the series, and probability distributions, used in computation of the redundancies $R$, are estimated as time-averaged histograms. When the discrete variables $X_1, \ldots, X_n$ are obtained from continuous variables on a continuous probability measure space, then the redundancies $R$ depend on a partition $\xi$ chosen to discretize the space. Various strategies have been proposed to define an optimal partition for estimating redundancies of continuous variables (see [17, 18] and references therein). Here we use the "marginal equiquantization" method described in [18, 20].

The marginal redundancy $\varrho(X(t), X(t+\tau_1), \ldots, X(t+\tau_{n-1}); X(t+\tau_n))$ reflects a specified model. Not always, however, a particular model can be evaluated. An order of a prediction model can be established in a fitting procedure and thus can be variable or unknown before the predictor is fitted. Also, computation of a redundancy of the type $\varrho(X(t), X(t+\tau_1), \ldots, X(t+\tau_{n-1}); X(t+\tau_n))$ means an estimation of $(n+1)$-dimensional probability distribution. This can be problematic, especially when short time series (or relatively short segments of a longer series) needs to be evaluated. In such cases the evaluation of simpler redundancies (and related lower dimensional probability distributions) can be more reliable, in particular, the evaluation of a simple mutual information $I(x(t); x(t+\tau_n))$, where $\tau_n$ is the prediction horizon, can be more reliable then the evaluation of high-dimensional redundancies of the type $\varrho(X(t), X(t+\tau_1), \ldots, X(t+\tau_{n-1}); X(t+\tau_n))$.

Whatever the dimension of the estimated probability distribution is, the estimates of the redundancies of continuous variables are always directly influenced by a partition chosen to discretize the continuous probability space. Then a result like "the theoretical predictability is 12.5 bits" is practically useless. Therefore we do not compute the theoretical predictability (given by a particular redundancy value) in its absolute values, but relative to the maximum redundancy possible in a particular situation. Considering the mutual information (4) and marginal redundancies of the type (7), the maximum value is given by the entropy of a series $\{x(t)\}$. If the method of marginal equiquantization is applied, it is equal to $\log Q$, where $Q$ is the number of marginal equiquantal boxes [18, 20]. In the following we normalize the values of the theoretical predictability by the entropy of the studied series and thus we obtain the theoretical predictability in per cents of a maximum predictability (a one-to-one map considering given precision).

---

[3]Multivariate nonlinearity tests and related surrogate data are discussed in [21, 22].

Estimating the theoretical predictability from real data, the predictability estimates are always positive and, having very noisy data, it is hard to evaluate actual difference of a data from an unpredictable white noise. Therefore it is desirable to compute an indicator related directly to this problem. We propose to use statistics used in the surrogate data tests: If any predictability is under question, compute a redundancy from the data under study and a set of the scrambled surrogates. Then compute a statistical quantity as a difference between the redundancy value obtained from the data and a mean value of a set of the scrambled surrogates, in standard deviations (SD's) of the latter (denoted as "NONLINEAR vs. IID" in the following). When nonlinear predictability, as opposed to predictability given by linear autocorrelation, is investigated, the same statistical quantity based on the isospectral (FT-based) surrogates should be computed ("NONLINEAR vs. LINEAR"). (See [20].) Evaluating specifically linear predictability, one can evaluate either the difference of the *linear* redundancies, introduced in Sec. 2, between the data and the scrambled surrogates, or the difference of the general redundancies $R$ between the scrambled surrogates and the isospectral surrogates ("LINEAR vs. IID"). As the output of the method one can obtain three values of the theoretical predictability (i.e., the redundancy in per cents of the maximum possible redundancy):

1. Nonlinear predictability, or, more precisely, total predictability, which includes both linear and nonlinear dependences, evaluated by the general (nonlinear) redundancies $R$ from the data under study.

2. Linear predictability, obtained either by the linear redundancies from the data under study, or by the general redundancies from the isospectral surrogates.

3. Numerical zero of the method, i.e., the theoretical predictability related to a white noise with the same mean, variance, histogram and number of samples as the data under study (estimated from the scrambled surrogates).

In addition, three types of statistical indicators, listed above ("NONLINEAR vs. IID", "NONLINEAR vs. LINEAR", and "LINEAR vs. IID") are evaluated, which indicate whether (and how much) the studied data are significantly different from a particular type of the surrogate data.

The time-averaged marginal redundancy $\varrho(x(t), x(t+\tau_1), \ldots, x(t+\tau_{n-1}); x(t+\tau_n))$ is, provided that the series $\{x(t)\}$ is stationary, independent of $t$ and dependent on $\tau_1, \ldots, \tau_n$. Then, the theoretical predictability and the related statistical quantities can be evaluated as functions of $\tau$'s and/or the number $n$ of a model inputs. Or, as a different task, a particular model, say $\tau_i = i\tau$, where $\tau$ is the sampling time, can be chosen and a series can be scanned for changes in predictability using overlapping windows ("a slicing window") of the length $N_w$ and the window step $N_s$. Both these approaches, considering the statistical quantities, yield multiplicity of test values and thus open a question of simultaneous statistical inference, which is discussed in [19, 20, 16, 13, 14, 8].

The linear redundancy, introduced in Sec. 2, can be used either to evaluate the linear predictability, as discussed above, or to check quality of the isospectral surrogates in order to avoid spurious detection of nonlinearity (nonlinear predictability) caused by imperfect isospectral surrogates. This problem is discussed in detail in [20] and will not be considered further in this paper, although we propose to check all results about nonlinearity (nonlinear predictability) using related linear redundancies, as described in [20].

# 5 Examples

We demonstrate the proposed method using one numerically generated time series with variable predictability and three series of real data from foreign exchange market (forex). In all cases we evaluate the simple mutual information $I(x(t); x(t+\tau))$, where the prediction horizon $\tau$ is equal to the sampling time (one-step ahead prediction problem). The predictability $I(x(t); x(t+\tau))$ and the related statistics are evaluated in overlapping windows with $N_w = 256$ and $N_s = 4$ or 8 samples. The analyzed data are clearly nonstationary, we employ a working hypothesis of step-wise stationarity (see [1] and references within), in other words, we suppose that breaking the requirement of stationarity is not so strong in individual windows as in the whole time series. The theoretical predictability and related statistics are compared with the actual varying predictability in the case of numerically generated series (varying predictability of which is given by the way of its construction), and with accuracy of a simple nonlinear (spline) predictor in the case of the forex data.

## 5.1 Artificial data

Consider a time series:
$$x(t) = a(t)\alpha(t) + b(t)\Lambda(t) + c(t)\sigma(t), \qquad (15)$$

where $\alpha(t)$ is a realization of a linear autoregressive process of the first order, $\Lambda(t)$ is a time series obtained by integrating the nonlinear (and chaotic) Lorenz system [12] and $\sigma(t)$ is Gaussian noise with zero mean and unit variance. The time dependent parameters $a(t)$ and $b(t)$, as functions of time, are displayed in Fig. 1. The parameter $c(t)$ is defined as $c = 1 - \max(a, b)$. The resulted time series $\{x(t)\}$ is displayed in Fig. 1, top panel.

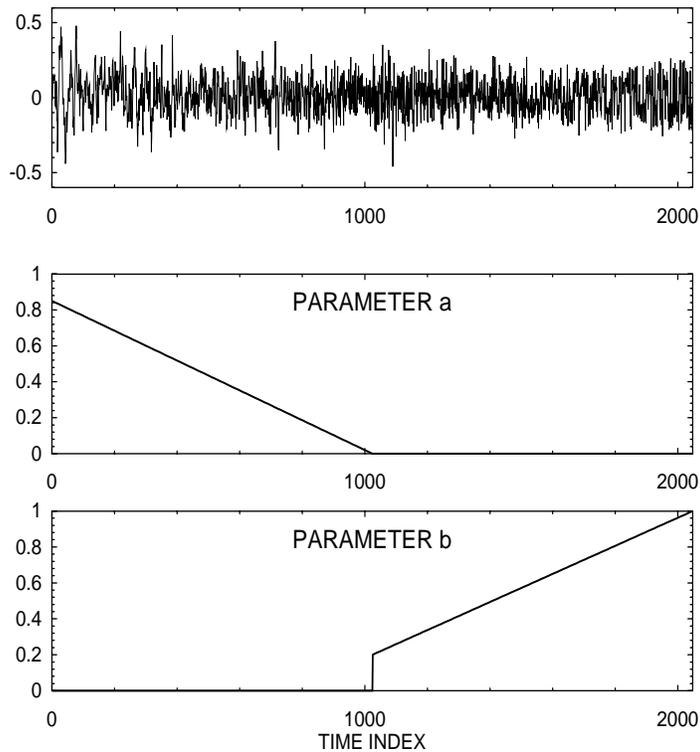

Figure 1: Numerically generated time series (top panel); time evolution of parameters $a(t)$ (middle panel) and $b(t)$ (bottom panel) used in generating the series.

The theoretical predictability and the related statistical indicators, computed from the mutual information $I(x(t); x(t+\tau))$ in a slicing window, are displayed in Fig. 2. One can see that both the general and linear predictability decrease in the first half of the series (and their difference is inside the error of the estimates), while in the second half of the series the nonlinear predictability increases sharply, while the linear predictability increases only slightly over the white noise level. This finding is consistent with the fact, that the first half of the series contains a data from a linear process, i.e., the linear predictability is the only (and total) predictability of the series, while, in its second half, the series contains the data from the nonlinear system with strong nonlinear but only weak linear temporal dependences. The portions of these processes to white noise in the series change according to the parameters $a$ and $b$ (Fig. 1). The relation between the actual "amount" of the predictable processes in the series (i.e., between the parameters $a$, $b$) and the theoretical predictability and the related statistics is not linear, but these indicators correctly capture the evolution of predictability of the series.

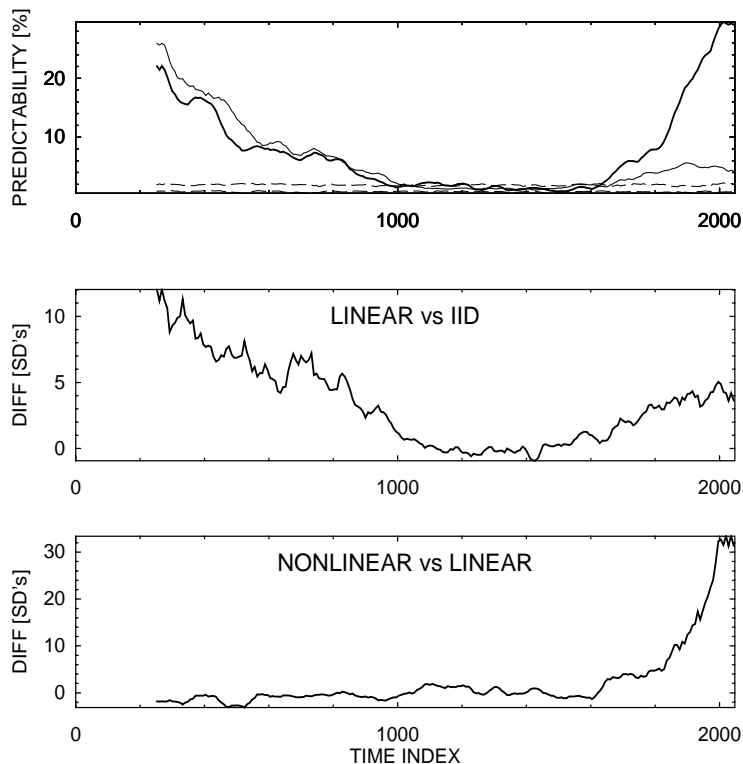

Figure 2: *Top panel: Theoretical predictability for the numerically generated series. Nonlinear (total) predictability (full thick line), linear predictability (full thin line) and predictability related to white noise (values of mean ± SD for the scrambled surrogates, dashed lines). Middle panel: Statistical indicator of linear predictability. Bottom panel: Statistical indicator of nonlinear predictability (as additive to linear predictability, i.e., results of tests against isospectral surrogates). All quantities were computed in the slicing window ($N_w = 256$, $N_s = 8$).*

## 5.2 Forex data

High-frequency foreign exchange data used here have been collected from the Reuters terminal. The series are not sampled in real time, but in so-called tick time. (Each value of the series is a median over 100 (the first example) or 50 (the other two examples) usual (log(bid)+log(ask))/2 values. The data were differenced prior to the analysis.) We conjecture that this is a "natural" time of underlying market dynamics and the method is applied as in the case of series sampled equidistantly in real time[4]. The theoretical predictability and the related statistics, again obtained from the mutual information $I(x(t); x(t+\tau))$ in a slicing window, are compared with prediction errors of a smoothing spline predictor, which is based on approximation of a polynomial smoothing spline with a continuous derivation of first and second order in different windows. The prediction is based on extrapolation of the spline outside of the window. For each time point a different window and different type of spline is used because the parameters were automatically optimized from last week data.

Results for 100-tick USD/DEM series are presented in Figs. 3 and 4. The prediction errors in each point (Fig. 3, top panel) are hard to compare with the indicators of the theoretical predictability. Therefore we have computed averaged absolute errors over the same slicing windows as the theoretical predictability ("average slicing errors", middle panel in Fig. 3). The theoretical predictability is presented in the bottom panel of Fig. 3. The extent of the method's "numerical zero", i.e., the values of mean ± SD for the scrambled surrogates, is displayed by dashed lines. Full lines are used for the linear (thin full line) and nonlinear (thick full line) predictability.

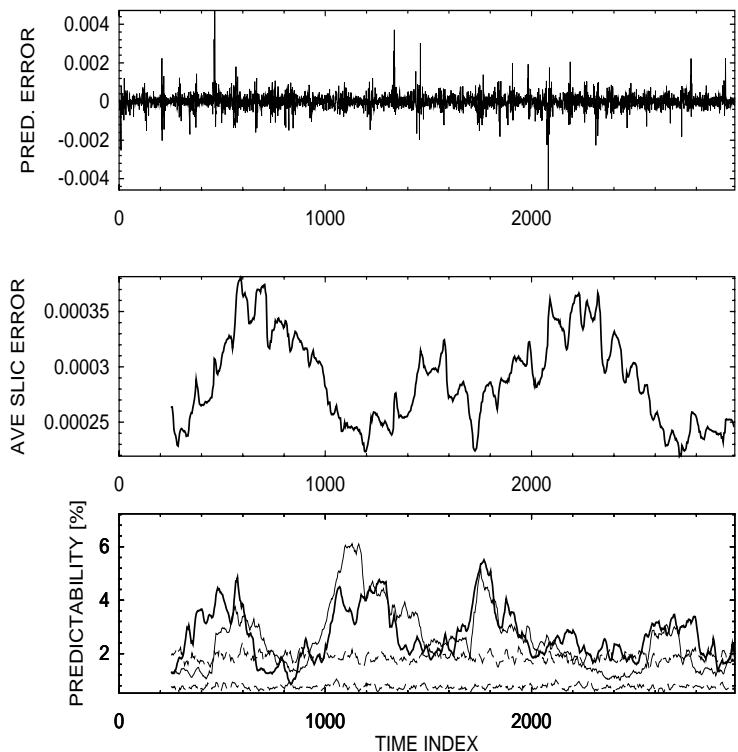

Figure 3: *Results for the 100-tick USD/DEM series. Top panel: Prediction errors of a spline predictor. Middle panel: Averaged slicing absolute errors, i.e., absolute averages of the prediction errors computed in the slicing window ($N_w = 256$, $N_s = 8$). Bottom panel: The theoretical predictability, computed in the slicing window ($N_w = 256$, $N_s = 8$). Nonlinear (total) predictability (thick full line), linear predictability (thin full line), predictability for white noise (values of mean ± SD for the scrambled surrogates, dashed lines).*

---

[4] The problem of time in economic time series is discussed in [4]. Detailed discussion in context of the proposed method and related redundancy/surrogate data nonlinearity tests will be published in near future.

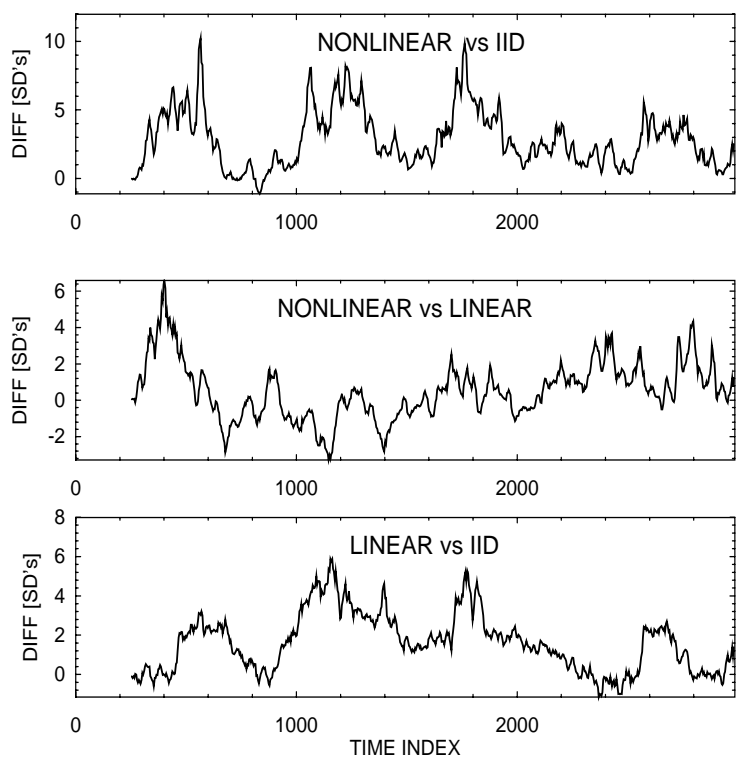

Figure 4: *Results for the 100-tick USD/DEM series. Statistical indicators of: nonlinear (total) predictability (top panel), nonlinear (as additive to linear) predictability (middle panel) and linear predictability (bottom panel). All quantities were computed in the slicing window ($N_w = 256$, $N_s = 4$).*

Statistical indicator for the total predictability ("NONLINEAR vs. IID") is displayed in Fig. 4, top panel. One can see that, with the exception of the beginning of the series, the minima of the theoretical predictability agree with the maxima of the averaged slicing error, and vice versa. The statistical indicator of nonlinearity ("NONLINEAR vs. LINEAR", Fig. 4, middle panel) shows that at the beginning of the series there is a nonlinear dependence, which was not utilized by the spline predictor. Comparing all the three statistical indicators in Fig. 4 one can conclude, that in this case the predictor profits from both linear and nonlinear structures (except of the segment at the beginning, mentioned above) and the best indicator of predictability, in this case, is the statistic related to the total predictability ("NONLINEAR vs. IID, Fig. 4, top panel).

Results for 50-tick USD/JPY series are presented in Fig. 5. The decreasing trend of the averaged slicing error (Fig. 5, top panel) between 500 and 1300 samples ("time index") can be compared to related increase in nonlinear and total predictability (Fig. 5, middle panel). The evolution of the linear predictability (Fig. 5, bottom panel) seems irrelevant. The opposite result was observed by processing a 50-tick GBP/USD series (Fig. 6): Here the prediction errors (Fig. 6, top panel) are consistent with the linear predictability indicator (Fig. 6, middle panel), but not with the indicators of the nonlinear and total predictability.

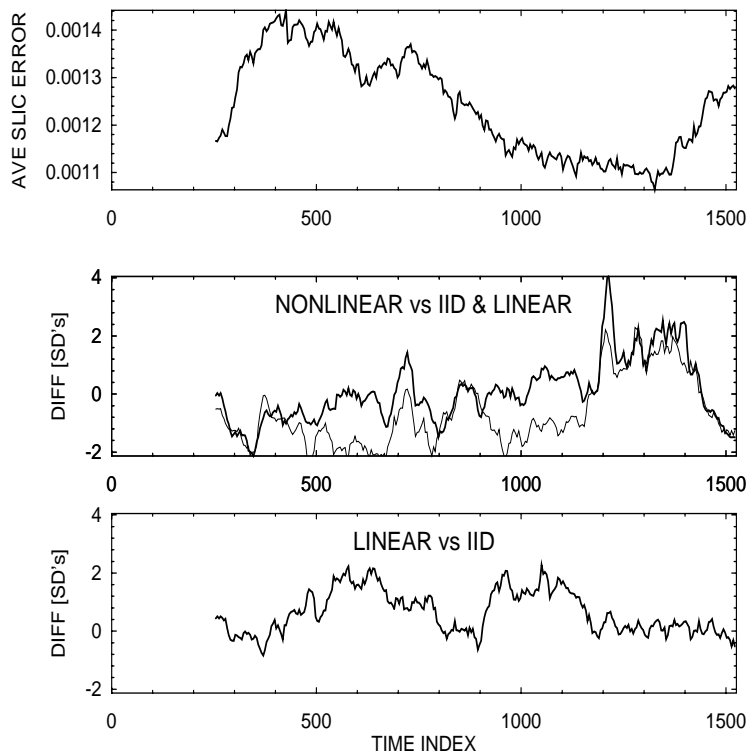

Figure 5: *Results for the 50-tick USD/JPY series: Averaged slicing absolute error (top panel), statistical indicators of nonlinear—total predictability (thick line) and nonlinear (as additive to linear) predictability (thin line, both middle panel), and linear predictability (bottom panel). All quantities were computed in the slicing window ($N_w = 256$, $N_s = 4$).*

## 6 Discussion and Conclusion

The method for estimating theoretical predictability of time series was presented. It is based on evaluating the information-theoretic functionals—redundancies, which quantify an amount of information between time series related to an input of a prediction model and a time series on the model output, i.e., the series lagged by the prediction horizon from the input series. As far as the estimated redundancy values depend on an estimation method and precision used, the resulted redundancies are normalized by a maximum possible redundancy value and presented as the theoretical predictability in per cents of maximum hypothetical predictability (a one-to-one map considering given precision). Further, the same theoretical predictability measures (redundancies) are evaluated from two types of surrogate data, in order to test statistically differences of the series under study from
a) white noise with the same mean, variance and histogram as the raw data (the scrambled surrogates), and
b) a linear stochastic process with the same spectrum as the raw data (the isospectral, or FT surrogates).
The series can be classified as either unpredictable (not different significantly from white noise), or predictable. Its predictability can be distinguished as linear or nonlinear, and proposed quantitative measures of predictability can be used for comparison either with some known properties of the studied data or with performance of a predictor.

The proposed method was demonstrated using both numerically generated and real data. In the case of the artificial data, which were generated to have variable predictability of two types (linear in the first, and nonlinear in the second half of the series), the type and the level of predictability were correctly indicated by the proposed method.

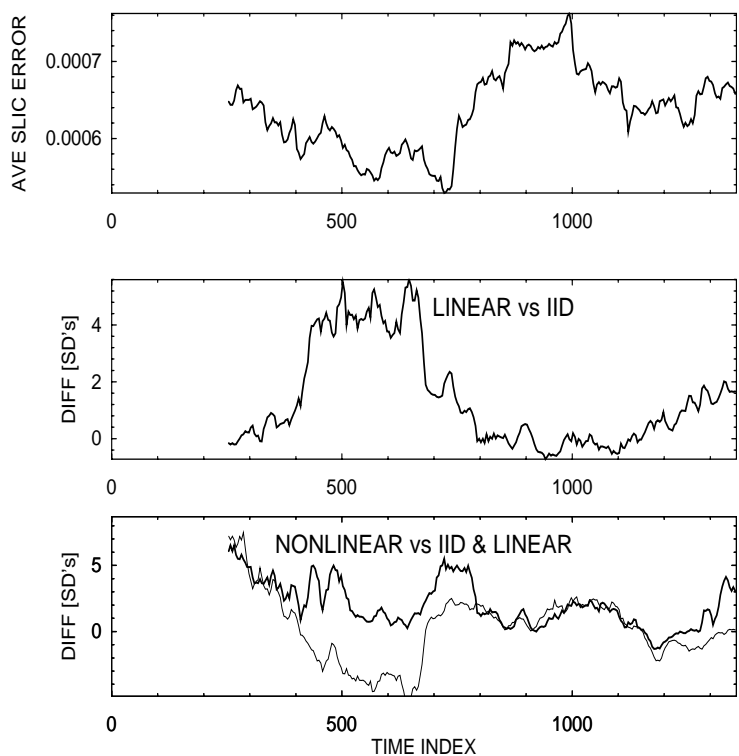

Figure 6: *Results for the 50-tick GBP/USD series: Averaged slicing absolute error (top panel), statistical indicators of linear predictability (middle panel) and nonlinear (total) predictability (thick line) and nonlinear predictability (thin line, both bottom panel). All quantities were computed in the slicing window ($N_w = 256$, $N_s = 4$).*

In the second example, real data—foreign exchange rates of three currencies (DEM, JPY, GBP) relative to USD were used and the proposed measures of the theoretical predictability were compared to prediction errors of a spline predictor. These results should be considered preliminary. One could see that there are several indicators of predictability and it was not clear which one is optimal. Further study is necessary, in which more data will be processed, employing also different kinds of predictors. As the aim of the study, an optimal indicator of the theoretical predictability will be chosen for a particular predictor, its critical values will be found and a marketing strategy will be proposed.

## Acknowledgements

Reuters and H.E.M. Informatics are acknowledged for providing the forex data. This study is supported by the Grant Agency of the Czech Republic (grant No. 201/94/1327) and by the Academy of Sciences of the Czech Republic (grant No. 230404).